\documentclass[journal,twoside,web]{ieeecolor}
\usepackage{generic}
\usepackage{cite}
\usepackage{amsmath,amssymb,amsfonts}
\usepackage{graphicx}
\usepackage{textcomp}
\usepackage{enumerate}
\usepackage{caption}
\usepackage{subcaption}
\usepackage{caption}
\usepackage{hyperref}
\usepackage{tikz}
\usepackage{dblfloatfix}
\usetikzlibrary{spy}
\usepackage{siunitx}
\DeclareMathOperator*{\argmin}{\arg\!\min}

\usepackage{algorithm}
\usepackage{algpseudocode}
\usepackage{xcolor}

\newcommand*\Let[2]{\State #1 $\gets$ #2}

\DeclareMathOperator{\pred}{\mathrm{pr}}
    
\newcommand{\bs}[1]{\boldsymbol{#1}}

\newcommand{\rev}[1]{{#1}}

\begin{document}

\title{Neural Network Kalman filtering for 3D object tracking from linear array ultrasound data}

\author{Arttu Arjas, Erwin J. Alles, Efthymios Maneas, Simon Arridge, Adrien Desjardins, Mikko J. Sillanp\"a\"a, and Andreas Hauptmann, \IEEEmembership{Member, IEEE} 
\thanks{Submitted for review on xyz. This work was supported by Academy of Finland Projects 336796, 326291, and 338408, the CMIC-EPSRC platform grant (EP/M020533/1), the Wellcome Trust (203145Z/16/Z), the Engineering and Physical Sciences Research Council (EPSRC) (NS/A000050/1), and the Rosetrees Trust (PGS19-2/10006)}
\thanks{A. Arjas and M. Sillanp\"a\"a
are with the Research Unit of Mathematical Sciences, University of Oulu, Finland.}
\thanks{E. Alles, E. Maneas and A Desjardins are with the Department of Medical Physics \& Biomedical Engineering, University College London, UK and with the Wellcome/EPSRC Centre for Interventional and Surgical Sciences, University College London, UK.}
\thanks{S. Arridge is with Department of Computer Science, University College London, UK.}
\thanks{A. Hauptmann is with the Research Unit of Mathematical Sciences, University of Oulu, Finland and with the Department of Computer Science, University College London, UK.}
}

\maketitle

\begin{abstract}
Many interventional surgical procedures rely on medical imaging to visualise and track instruments. Such imaging methods not only need to be real-time capable, but also provide accurate and robust positional information. In ultrasound applications, typically only two-dimensional data from a linear array are available, and as such obtaining accurate positional estimation in three dimensions is non-trivial. In this work, we first train a neural network, using realistic synthetic training data, to estimate the out-of-plane offset of an object with the associated axial aberration in the reconstructed ultrasound image. The obtained estimate is then combined with a Kalman filtering approach that utilises positioning estimates obtained in previous time-frames to improve localisation robustness and reduce the impact of measurement noise. The accuracy of the proposed method is evaluated using simulations, and its practical applicability is demonstrated on experimental data obtained using a novel optical ultrasound imaging setup. Accurate and robust positional information is provided in real-time. Axial and lateral coordinates for out-of-plane objects are estimated with a mean error of 0.1mm for simulated data and a mean error of 0.2mm for experimental data. Three-dimensional localisation is most accurate for elevational distances larger than 1mm, with a maximum distance of 6mm considered for a 25mm aperture.
\end{abstract}

\begin{IEEEkeywords}
Kalman filtering, neural networks, object tracking, out-of-plane artefacts, optical ultrasound
\end{IEEEkeywords}

\section{Introduction}

Tracking and localisation of point-like objects is crucial for a large variety of medical applications in ultrasound (US) imaging, such as tracking of microbubbles for super-resolution US imaging \cite{ackermann2015detection,christensen2020super} or US-guided placement of fiducial markers for radiotherapy \cite{dhadham2016endoscopic}. Additionally, tracking of surgical tools (such as needles and catheters) is essential during minimally invasive procedures \cite{chin2008needle,helbich2004stereotactic,daffos1985fetal}, as when placed inaccurately, these devices may cause trauma by damaging tissue or deliver ineffective treatment to the wrong location \cite{karakitsos2006real,daffos1985fetal}. 
As such, US is frequently used for guidance through imaging but accurate localisation in a three-dimensional target domain remains challenging. 
This is primarily caused by the nature of data acquisition using linear arrays, which assumes that all signals originate from within the image plane and thus only a two-dimensional B-mode image of the image plane is formed. We refer to this obtained 2D image as the US image and assume it consists of the reconstructed point, or point-like, source corresponding to the object we aim to track accurately.
But, if this point source is located out-of-plane it will primarily show as aberration in the reconstructed image domain; additionally one may misinterpret features, such as a needle shaft as the tip \cite{ultrasoundtracking}. Thus, the problem to provide an accurate positional estimate in 3D from only 2D US images is consequently a notoriously difficult task without any auxiliary information \cite{beigi2020enhancement} and is a field of active research \cite{guo2014active,graham2019vivo}. Early approaches used speckle information to estimate out-of-plane displacements\cite{krupa2007real,afsham2013generalized}. Another possibility for instrumented US tracking of needles was proposed by Xia et al.  \cite{xia2017looking,3Dultrasound}, who designed a custom-made imaging probe consisting of a central array for conventional imaging and two side arrays for 3D tracking \cite{3Dultrasound}. Alternatively, one may approach the needle tracking problem in full 3D to obtain accurate positional estimates \cite{3DUSkalman}.

\begin{figure*}
    \centering
    \includegraphics[width=0.8\textwidth]{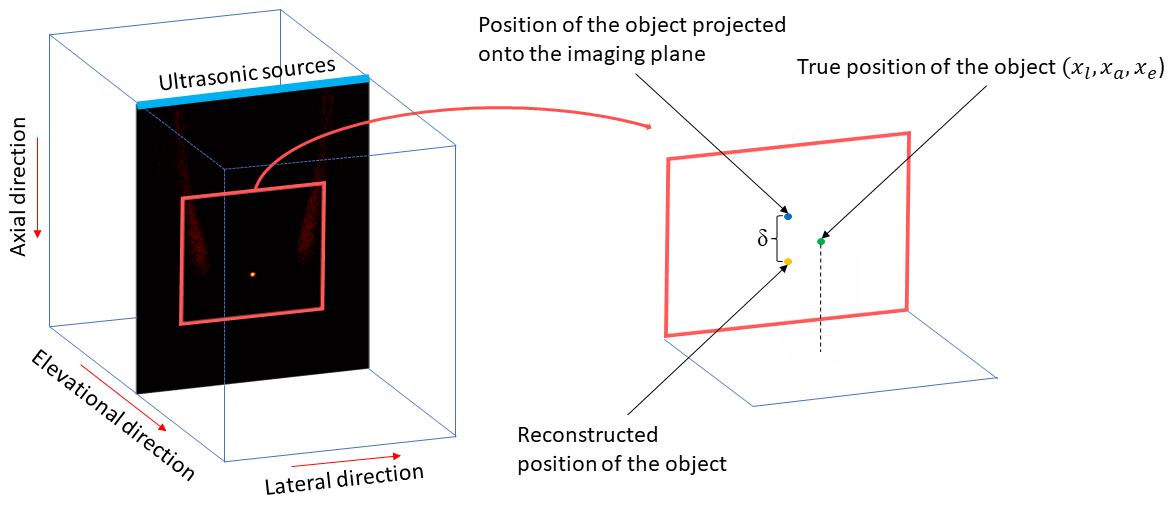}
    \caption{Illustration of the measurement geometry. If the point source object is located out-of-plane (right side), then the object appears distorted in the reconstructed US image, i.e. too low by $\delta$ in the axial direction. This is why the axial position needs to be corrected.}
    \label{fig:geometry}
\end{figure*}

In this work, we propose an alternative, real-time capable, method of performing 3D tracking without the need for custom-made probes and using a single set of measurements per time-step from a linear array. In the following we assume a point source model for the tracked object.
For the estimation of lateral and axial positions, we examine high intensity pixels in the reconstructed 2D US image, similar to \cite{3DUSkalman}, where tracking was performed in full 3D. For the estimation of the elevational direction, or out-of-plane distance of the point source to the image plane, we use a machine learning approach. In particular, significant markers are extracted from the measured time series and a neural network is trained using synthetic data modelled for a prototype optical ultrasound (OpUS) imaging setup  \cite{alles2021freehand} 
to predict out-of-plane distance and associated aberration in axial position in the reconstructed 2D US image. The markers used for the neural network are summary statistics extracted from the measurement data and correlated with the offset to establish a nonlinear mapping between the two. 

Additionally, we assume regularity in the temporal evolution of the object location to improve robustness and reduce uncertainty in the estimation compared to independently analysing subsequent images. The regularity assumption mimics conditions encountered in clinical practice, where the objects, such as needle tips and microbubbles, are expected to follow a smooth trajectory without rapid jumps or jitter during insertions into soft tissue. This can be incorporated, while retaining computational efficiency, by a Kalman filter \cite{kalman1960new,bayesfilter} which is a flexible method for estimating the state (position, velocity, etc.) of a dynamic system. 
It has been classically utilised in engineering applications such as target tracking and navigation, but has been also extensively used in inverse problems and medical imaging \cite{kaipio2006statistical,prince2003time, liang2020nonstationary,3Dultrasound,bayesfilter,hakkarainen2019undersampled}. The underlying idea of Kalman filtering is to update the estimate of the state at time step $k + 1$ each time new data become available as opposed to smoothing, where the whole trajectory from time step $1$ to $k$ are updated as well. It has the appealing property, as opposed to estimating the full posterior of all states, that the problem does not become intractable as the number of data points increases.

We note, that Kalman filtering has earlier been utilised for a needle tracking problem in 3D \cite{3DUSkalman} as well as for microbubble tracking in 2D \cite{solomon2019exploiting}. In \cite{takeshima2021position}, the authors track a wire tip using Kalman filter and perform an elevational position estimation from geometric markers.
In this work we approach the problem in 3D with only a single set of measurements (per time point) from a linear array and combine it with a neural network in order to obtain reliable estimates on the elevation to correct the axial position in the 2D US image $\hat{x}$. 
We evaluate the proposed method by tracking a point-source 
for simulated OpUS data and object trajectories with changing elevation. Robustness is evaluated with respect to increased noise in the measurement data and accuracy compared to positional estimation using only the pixel with maximum intensity in the OpUS image. Finally, we evaluate the method on experimental OpUS measurements. 

\section{A Kalman filter for object tracking and out-of-plane correction}
\subsection{Image formation and object tracking}
In the following we specifically consider a custom setup and  simulation framework for a freehand optical ultrasound imaging system. That is, the US signal generation is modelled as pulse-echo imaging, where each source along the linear aperture emits a pressure wave, which reflects off the point scatterer and is detected by a single fibre-optic detector placed right next to the imaging aperture \cite{alles2021freehand}. We note that the tracking framework here can be generalised to any linear US array, where every source element also acts as a detector. Moreover, we consider in the following the tracking problem in a 3D space, that is we aim to determine the 3D coordinate $x$ of the point-source.
Given the recorded radiofrequency (RF) time series $p$, reconstruction of the 2D in-plane US image $\hat{x}$ is performed using a basic delay-and-sum algorithm (equivalent to dynamic focussing) \cite{jensen2006synthetic}.

When the location of the point-source is in-plane, then the reconstructed image $\hat{x}$ can be directly used to estimate the location $x$ reliably. On the other hand, if $x$ is out-of-plane then the 2D reconstruction will lead to a distorted image, in the sense that the reconstructed axial position is located deeper in the target than the correct position \cite{cobbold2006foundations}. The aberration occurs because the time of flight is larger from objects that are positioned out-of-plane due to the imaging geometry (see Fig. \ref{fig:geometry}).
Thus, this axial aberration needs to be detected and processed to simultaneously provide an estimate of the elevational distance between the point target and the image plane, and the corresponding coordinates projected onto the image plane.

\subsection{Object tracking for in-plane objects}
Most tracking applications primarily assume that the object of interest is in-plane and features extracted from the reconstructed image $\hat{x}$ are good indicators of the actual position $x$. 
Thus, the majority of tracking algorithms are based on intensity values in the B-mode images for point marker tracking \cite{ipsen2016online,de20152014} in combination with various image registration approaches \cite{harris2010speckle,henriques2014high,vercauteren2008symmetric}. More recent developments make use of deep learning techniques to estimate the coordinates of objects directly from the measured time series $p$ \cite{allman2018photoacoustic,allman2019deep,yazdani2021simultaneous}.
Nevertheless, there is no clear gold-standard to perform object tracking, as the particular approach depends heavily on the application and practical need \cite{beigi2020enhancement}.

In this study we are concentrating on single object tracking and use the maximum intensity (MI) estimate for comparison and reference, since it provides highly efficient and accurate estimates under ideal assumptions, i.e., high signal-to-noise ratio without any elevation. Consequently, we will also design our tracking model in the following by using the pixels in the US image with highest intensity for the estimation of axial and lateral positions in the filtering process.

\begin{figure*}
    \centering
    \includegraphics[width=0.9\textwidth]{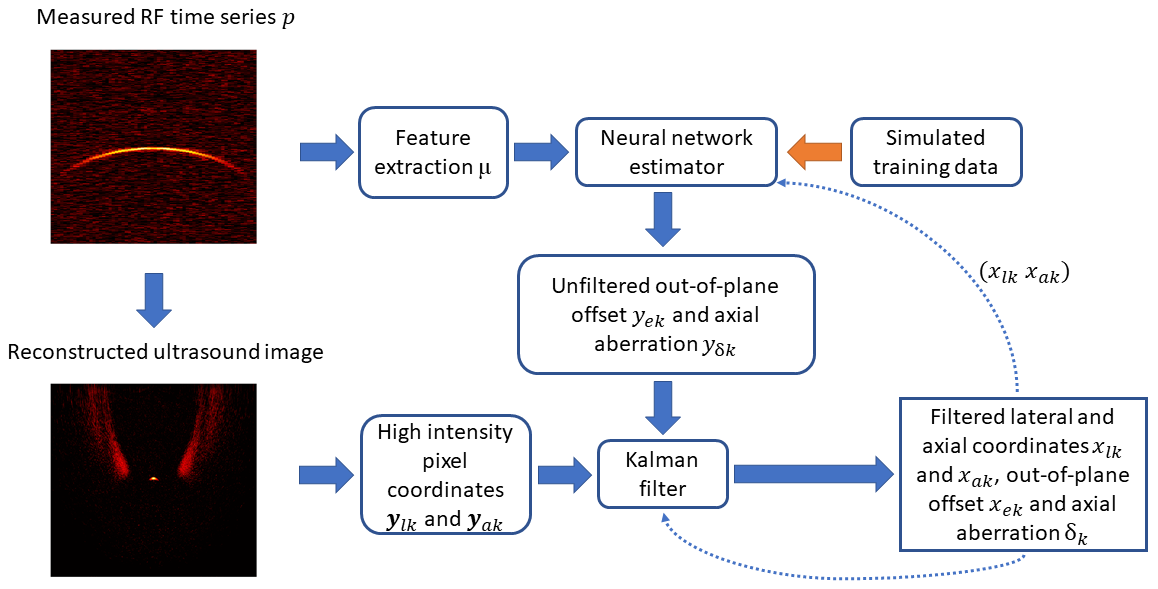}
    \caption{Flowchart for the Neural Network Kalman (NNK) filtered tracking. We extract mean value $\mu$ from the highest amplitude entries in the measured time series $p$ and use highest intensity pixels in the 2D US image. A previously trained (orange arrow) neural network estimator then uses the last state estimates of lateral $x_{lk}$ and axial $x_{ak}$ coordinates together with $\mu$ to estimate offset and axial aberration. The next state is then updated via Kalman filtering to provide robust positional estimation.}
    \label{fig:flowchart}
\end{figure*}

\subsection{Kalman filtering} \label{sec:kalmanfilter}
Kalman filtering, a class of Bayesian filtering, is especially effective in situations where \rev{the data stream is over time} and one must update the state given the new data and the history of the system; as such it is ideally suited to robustly perform the object tracking considered in this study. Specifically, Kalman filtering \cite{kalman1960new} consists of closed-form update formulas for a linear Gaussian filtering problem, which will be discussed next.
The estimation of axial and lateral coordinates is similar to the approaches suggested by  \cite{3DUSkalman,solomon2019exploiting}, we will then continue to extend our model to incorporate elevation and a correction of estimated axial coordinates.

\subsubsection{Lateral and axial coordinates}
Estimation of lateral and axial coordinates and corresponding velocities $\bs{x}_k = (x_{lk} \ x_{ak} \ v_{lk} \ v_{ak})^\mathsf{T}$ at time step $k$ is based on the locations of highest absolute intensity pixels in the image. We assume that these locations are spread around the location of the object. While the velocity of the object is not the main interest, it is introduced as an auxiliary variable to help in predicting the motion, as will be described below. We denote by $\bs{y}_{lk} \in \mathbb{R}^n$ the $n$ lateral and by $\bs{y}_{ak} \in \mathbb{R}^n$ the $n$ axial highest intensity locations and let $\bs{y}_k = (\bs{y}_{lk}^\mathsf{T} \ \bs{y}_{ak}^\mathsf{T})^\mathsf{T}$. We then build a model
\begin{equation}
    \bs{y}_k = \bs{Hx}_k + \bs{r}_k,
\end{equation}
where the matrix
\begin{equation}
\bs{H} = 
\begin{pmatrix}
1& 0 & 0 & 0 \\
\vdots& \vdots & \vdots & \vdots \\
1& 0 & \vdots & \vdots \\
 0 &1& \vdots & \vdots \\
 \vdots &\vdots& \vdots & \vdots \\
 0 &1& 0 & 0 
\end{pmatrix}
\in \mathbb{R}^{2n \times 4}
\end{equation}
associates given high intensity locations in $\bs{y}_k$ with the actual coordinates of the object in $\bs{x}_k$.

Furthermore, we assume that the noise in the observed locations is normally distributed $\bs{r}_k \sim \mathcal{N}(\bs{0}, \bs{R}_k)$, where 
\begin{equation}
\bs{R}_k = 
\begin{pmatrix}
s^2_{lk} \bs{I}_{n \times n}&\bs{0}\\
\bs{0}&s^2_{ak} \bs{I}_{n \times n}
\end{pmatrix}
\end{equation}
with $s^2_{lk}$ being the sample variance of $\bs{y}_{lk}$ and $s^2_{ak}$ the sample variance of $\bs{y}_{ak}$. This way uncertainty is naturally incorporated into the model as how spread out the high intensity locations are.

We model the motion of the object with a constant velocity model \cite{hong1998constantvelocity}
\begin{equation} \label{eq:dynamicmodel}
    \bs{x}_k = \bs{Ax}_{k-1} + \bs{Gc}_k,
\end{equation}
where $\bs{c}_k \sim \mathcal{N}(\bs{0}, \text{diag}(\sigma_l^2, \sigma_a^2))$ is assumed to be a random acceleration component and $\sigma^2_l$ and $\sigma^2_a$ are lateral and axial process noise variances, respectively. The matrices $\bs{A}$ and $\bs{G}$ are defined as \cite{bayesfilter}

\begin{equation}
    \bs{A} = 
\begin{pmatrix}
1& 0 &\Delta t&  0 \\
 0 &1& 0 &\Delta t\\
 0 & 0 &1& 0 \\
 0 & 0 & 0 &1
\end{pmatrix},
\end{equation}


\begin{equation}
\bs{G} = 
\begin{pmatrix}
\frac{1}{2}\Delta t^2 & 0\\
0 & \frac{1}{2}\Delta t^2\\
\Delta t & 0\\
0 & \Delta t
\end{pmatrix},
\end{equation}
where $\Delta t$ is the time between subsequent observations. 

The velocity and acceleration of the object at time step $k - 1$ is used to give an accurate prediction of the position at time step $k$. Writing Eq. \eqref{eq:dynamicmodel} explicitly for the positional variables only, we obtain
\begin{equation}
\begin{split}
    x_{lk} &= x_{l(k-1)} + \Delta t v_{l(k-1)} + \frac{1}{2}\Delta t^2 c_{lk}, \\
    x_{ak} &= x_{a(k-1)} + \Delta t v_{a(k-1)} + \frac{1}{2}\Delta t^2 c_{ak}.
    \end{split}
\end{equation}
This means that we assume the position at time step $k$ to be close to the position at time step $k - 1$ plus the displacement given by the time between subsequent observations, velocity, and acceleration.

\subsubsection{Out-of-plane offset and axial aberration}
In case there is an offset between the imaging plane and the object, we observe aberration in the reconstructed axial coordinate due to the geometry of the imaging problem (see Fig. \ref{fig:geometry}). In this case, location estimation based on only the high intensity pixels would result in a biased estimate of the axial coordinate. Instead, we use information in the measurement data $p$ to estimate the offset and axial aberration and use the knowledge to correct the estimate of the axial coordinate. To do this, we train a neural network with training data obtained from an OpUS simulator. Details on the neural network will be provided in following Section \ref{sec:neuralnet} and on the simulator in \ref{sec:simulator}. We note that instead of a neural network, other sufficiently expressive nonlinear prediction models could be used.

The filtering model is then extended to include the unfiltered out-of-plane offset and axial aberration, denoted as $y_{ek}$ and $y_{\delta k}$, that are received as output from the neural network. We can then define $\bs{y}_k^* = (\bs{y}_k^\mathsf{T} \ y_{ek} \ y_{\delta k})^\mathsf{T}$. Filtered out-of-plane offset and axial aberration and their velocities, denoted as $x_{ek}$, $\delta_{k}$, $v_{ek}$ and $v_{\delta k}$ are included in the state vector $\bs{x}_k^* = (\bs{x}_k^\mathsf{T} \ x_{ek} \ \delta_{k} \ v_{ek} \ v_{\delta k})^\mathsf{T}$, The extended model is
\begin{equation}
    \begin{split}
    \bs{y}_k^* &= \bs{H}^*\bs{x}_k^* + \bs{r}_k^*,\\
    \bs{x}_k^* &= \bs{A}^*\bs{x}_{k-1}^* + \bs{G}^*\bs{c}_k^*,
    \end{split}
\end{equation}
where $\bs{r}_k^* \sim \mathcal{N}(\bs{0}, \bs{R}_k^*)$, $\bs{c}_k^* \sim \mathcal{N}(\bs{0}, \text{diag}(\sigma^2_l,\sigma^2_a,\sigma^2_e,\sigma^2_\delta)),$
\begin{equation}
\bs{H}^* = 
\begin{pmatrix}
\bs{H} & \bs{0} \\
\bs{0} & \Tilde{\bs{H}}
\end{pmatrix},
\end{equation}
with
\begin{equation}
\Tilde{\bs{H}} = 
\begin{pmatrix}
1 & 0 & 0 & 0 \\
0 & 1 & 0 & 0
\end{pmatrix},
\end{equation}
\begin{equation}
\bs{A}^* = 
\begin{pmatrix}
\bs{A} & \bs{0} \\
\bs{0} & \bs{A}
\end{pmatrix},
\end{equation}
\begin{equation}
\bs{G}^* = 
\begin{pmatrix}
\bs{G} & \bs{0} \\
\bs{0} & \bs{G}
\end{pmatrix},
\end{equation}
and $\sigma^2_e$ and $\sigma^2_\delta$ are the process noise variances of the out-of-plane offset and axial aberration components, respectively. Finally, we let

\begin{equation}
\bs{R}_k^* = 
\begin{pmatrix}
\bs{R}_k & \bs{0} \\
\bs{0} & \Tilde{\bs{R}}_k
\end{pmatrix}
\end{equation}
where
\begin{equation}
\Tilde{\bs{R}}_k = \max(s_{lk}^2, s_{ak}^2) \bs{I}_{2 \times 2}.
\end{equation}

\subsubsection{State estimation}
As stated earlier, a major benefit arising from linearity and Gaussianity of the filtering models are the closed-form update formulas for mean $\bs{m}_k$ and covariance $\bs{P}_k$ of the state. At $k = 0$ we assume $\bs{x}_0^* \sim \mathcal{N}(\bs{m}_0, \bs{P}_0)$, where $\bs{m}_0 = \bs{0}$ and $\bs{P}_0 = 15\bs{I}$ to serve as an uninformative prior. Note, that the mean $\bs{m}_k$ corresponds to the estimated coordinates for $\bs{x}_k^*$ and $\bs{P}_k$ is the corresponding covariance matrix. At every round, the prior predictions for mean and covariance are updated recursively as
\begin{equation} \label{eq:kalmanupdate_pred}
    \begin{split}
        \bs{m}_k^{\pred} &= \bs{A}^*\bs{m}_{k-1},\\
        \bs{P}_k^{\pred} &= \bs{A}^*\bs{P}_{k-1}\bs{A}^{*\mathsf{T}} + \bs{Q}^*,
    \end{split}
\end{equation}
where $\bs{Q}^* = \bs{G}^*\text{diag}(\sigma^2_l,\sigma^2_a,\sigma^2_e,\sigma^2_\delta)\bs{G}^{*\mathsf{T}}$. We then evaluate the neural network as described in the following section to provide the estimates of offset and axial aberration in $\bs{y}^*_k$, then the Kalman update can be performed by
\begin{equation} \label{eq:kalmanupdate}
    \begin{split}
        \bs{u}_k &= \bs{y}_k^* - \bs{H}^*\bs{m}_k^{\pred}, \ \ \ \ \ \ \ \text{(Prediction residual)} \\
        \bs{S}_k &= \bs{H}^*\bs{P}_k^{\pred}\bs{H}^{*\mathsf{T}} + \bs{R}_k^*, \ \text{(Measurement covariance update)} \\
        \bs{K}_k &= \bs{P}_k^{\pred}\bs{H}^{*\mathsf{T}}\bs{S}_k^{-1}, \ \ \ \ \ \ \ \hspace{.5mm} \text{(Gain update)} \\
        \bs{m}_k &= \bs{m}_k^{\pred} + \bs{K}_k\bs{u}_k, \ \ \ \ \ \ \hspace{1mm} \text{(State update)} \\
        \bs{P}_k &= \bs{P}_k^{\pred} - \bs{K}_k\bs{S}_k\bs{K}_k^{\mathsf{T}}. \ \ \  \text{(State covariance update)}
    \end{split}
\end{equation}
Estimates of all coordinates are then given by $\bs{m}_k$. Their variances can be found in the diagonal of $\bs{P}_k$ and could be used for uncertainty quantification of the Kalman updates. The estimated axial aberration is then subtracted from the estimated axial coordinate to yield an estimate for the actual axial coordinate as
\begin{equation} \label{eq:distortioncorrection}
    m_{ak}^* = m_{ak} - m_{\delta k}.
\end{equation}

\begin{figure*}[t]
\centering
\begin{subfigure}{.5\textwidth}
  \centering
  \includegraphics[scale=0.35]{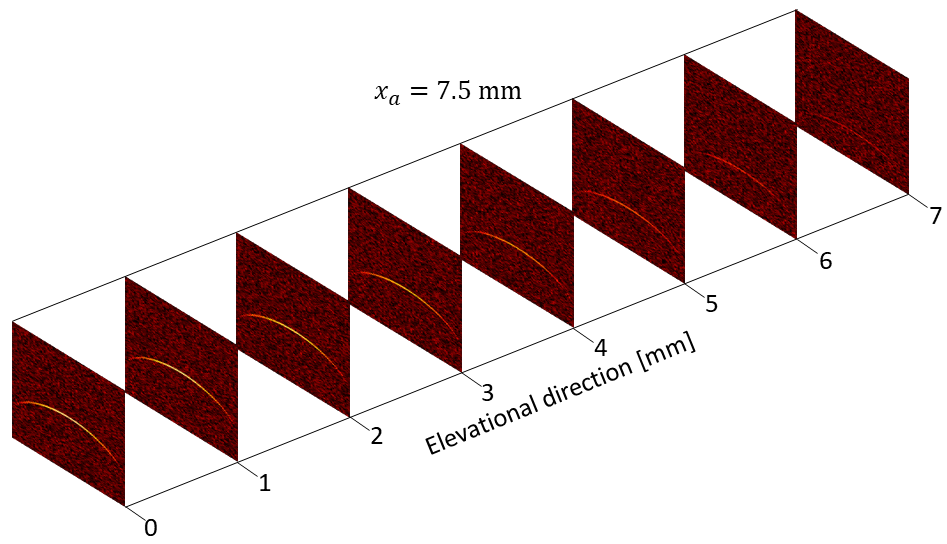}
\end{subfigure}%
\begin{subfigure}{.5\textwidth}
  \centering
  \includegraphics[scale=0.22]{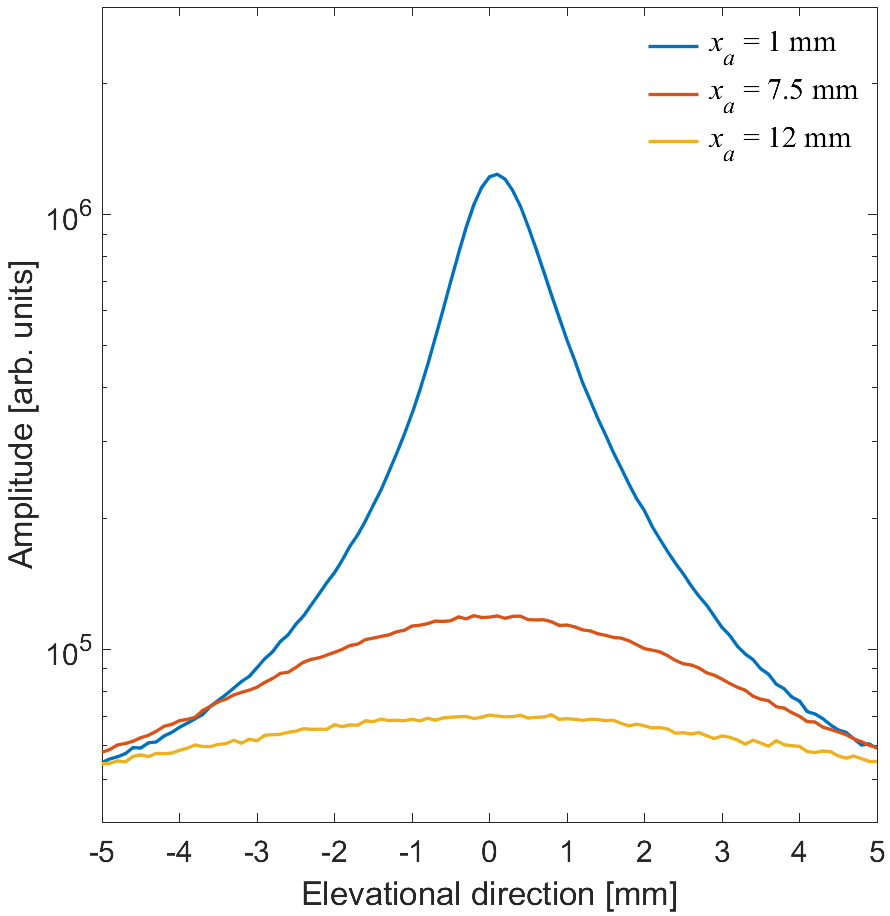}
\end{subfigure}
\caption{RF time series (left) showing the decay in amplitude as the distance from imaging plane increases. The decay is also shown on the right for different axial depths. In general, the rate of decay decreases with increased depth. The lateral coordinate $x_l$ was set to 0 in these tests.}
\label{fig:amplitude}
\end{figure*}


\subsection{Application to object tracking}\label{section:application}
We apply the Kalman filtering method to the task of object tracking modelled from OpUS image reconstructions and measurement data. 
After generating suitable training data and training the neural network, tracking can be performed as outlined in the following.

\subsubsection{Neural network: training data and architecture}\label{sec:neuralnet}
We use an OpUS simulator \cite{alles2018video,alles2020source} to generate training data for the neural network. First, we define a uniform $20 \times 20 \times 20$ grid of coordinates. The grid has bounds $\pm 12$ mm in lateral, $0.5$ -- $14.5$ mm in axial and \rev{$\pm 10$} mm in elevational direction. Two additional grid points were placed in-plane (zero in elevational direction). In each grid point we simulate measurement data with coordinate values equal to the grid point.

To obtain a marker for the offset estimation, we note that
ultrasound source elements typically emit near-omnidirectional pressure fields within the image plane, but are usually designed to emit highly directional fields in the out-of-plane direction. This is achieved through a combination of eccentric element geometries and acoustic lenses \cite{cobbold2006foundations}. As a result, the amplitude of pulse-echo signals from point objects depend strongly on the elevational (out-of-plane) position, and generally reduces with increasing elevational offset.
The shape of the decay also depends on the position of the object. In short, the out-of-plane amplitude decay decreases as the axial depth increases, as illustrated in Fig. \ref{fig:amplitude}. This is why both the lateral and axial coordinates are used as inputs for the neural network. Thus, the pulse-echo signal strength across the aperture can be used as an effective marker of the elevational position.
To exploit this, 
we use the RF time series $p$ to compute the mean absolute value $\mu$ of those time series samples belonging to either highest or lowest 1\% of a Gaussian defined by the mean and variance of $p$.
This way most of the purely noisy part of the data is ignored. We also compute the distance between the mean of apparent (reconstructed) axial coordinates of $n = 15$ highest intensity pixels and the real axial coordinate used to simulate the data. This distance reflects the axial aberration that needs to be corrected for.

A neural network $\Lambda_\theta$ with parameters $\theta$ is then trained to map lateral and axial coordinates, and mean absolute value of high amplitude entries in measured time series, denoted as $\bs{u} = (x_a \ x_l \ \mu)^\mathsf{T}$, to a prediction of unfiltered out-of-plane offset and axial aberration $\bs{w} = (y_e \ y_\delta)^\mathsf{T}$. 
Since the simulator output is almost symmetric with positive and negative offsets, we train the network with absolute offset values $\geq0$. This means that we can only estimate magnitude of the offset, not the direction. The network chosen is a standard multilayer perceptron \cite{goodfellow2016deep} with two hidden layers and 20 nodes in each layer. Each hidden layer has a sigmoid activation function whereas for the output layer the activation function is linear. The network is trained by finding a set of parameters $\theta^*$ such that mean squared error between neural network output and the ground truth is minimised, i.e.,
\begin{equation}
    \theta^* = \argmin_\theta \sum_{i=1}^M \|\Lambda_\theta(\bs{u}_i) - \bs{w}_i\|_2^2,
\end{equation}
where $M$ is the size and $i$ is an index over the training data. \rev{We used the Levenberg-Marquardt algorithm \cite{levenberg, marquardt} to train the neural network. The dampening parameter was set to the default value of $10^{-3}$. The optimisation stopped when the validation performance did not improve in six epochs in a row or the relative norm of the gradient of the minimised function was smaller than $10^{-7}$.}
\subsubsection{Tracking}
We track the point-source from a sequence of optical ultrasound image reconstructions. At $k = 1$, we set $\bs{m}_1 = \bs{0}$ and $\bs{P}_1 = 15\bs{I}$. We find the coordinates of $n = 15$ highest intensity pixels and use the neural network to estimate the unfiltered out-of-plane offset and axial aberration. Since the neural network input contains the lateral and axial position of the point-source, we use the estimate from the previous time step. 
If the motion of the object is somewhat regular, this does not have a big impact on the estimation accuracy. The coordinate estimates are then updated with Kalman filter update formulas Eq. \eqref{eq:kalmanupdate}. An estimate of the true axial coordinate of the object is then obtained by subtracting the axial aberration estimate from the apparent axial coordinate obtained directly from the 2D US image. An illustration of the full tracking workflow is shown in Figure \ref{fig:flowchart} and summarised as pseudo code in Algorithm \ref{alg:NNK}.
\begin{algorithm}
	\caption{NNK: Neural Network Kalman filtered tracking}
	\label{alg:NNK}
	\begin{algorithmic}[1]
	\State Initialisations: $\bs{m}_0 = \bs{0}$, $\bs{P}_0 = 15\bs{I}$
        \Function{NNK}{Inputs: process noise variances $\sigma_l^2$, $\sigma_a^2$, $\sigma_e^2$ and $\sigma_\delta^2$}
        \Let{$k$}{$1$}
        \While{new data acquired} 
        \State Update mean $\bs{m}_k^{\pred}$ and covariance $\bs{P}_k^{\pred}$ by Eq. \eqref{eq:kalmanupdate_pred}
        \State Compute marker $\mu_k$ and high intensity pixel locations $\bs{y}_{lk}$ and $\bs{y}_{ak}$
        \Let{$\bs{u}$}{$(m_{l(k-1)},m_{a(k-1)}^*,\mu_k)$}
        \Let{$(y_{ek} , y_{\delta k})$}{$\Lambda_\theta(\bs{u})$}
        \State Perform Kalman update with \eqref{eq:kalmanupdate}
        \State Perform axial aberration correction with \eqref{eq:distortioncorrection}
        \State Display image overlaid with coordinate estimates
        \Let{$k$}{$k+1$}
        \EndWhile
        \EndFunction
	\end{algorithmic}
\end{algorithm}

\section{Optical ultrasound and experiments}

\subsection{Experimental setup}
Experimental validation of the method was performed using a custom OpUS imaging system comprising a handheld imaging probe. We have chosen the OpUS imaging system for this study due to three main advantages: it offers direct access to the RF data, it was previously accurately characterised in-house, and the system can be accurately and highly efficiently modelled numerically -- thus making it an ideal fit for this study.
This system, which was described in full in \cite{alles2021freehand}, uses scanning optics to couple excitation light sequentially into the proximal ends of 64 optical fibres arranged in a linear array. This light is delivered to an optically absorbing coating deposited at the distal ends, where it is converted into divergent ultrasound waves via the photoacoustic effect \cite{beard2011biomedical}. Thus, an OpUS source aperture is rapidly scanned to enable video-rate and real-time imaging in a 2D imaging plane. Back-scattered ultrasound waves are detected using a single fibre-optic ultrasound detector comprising an optically resonant plano-concave Fabry-Pérot cavity \cite{guggenheim2017ultrasensitive}, with lateral extent of 25mm.


\begin{figure*}[!th]
    \centering
    \includegraphics[width=\textwidth]{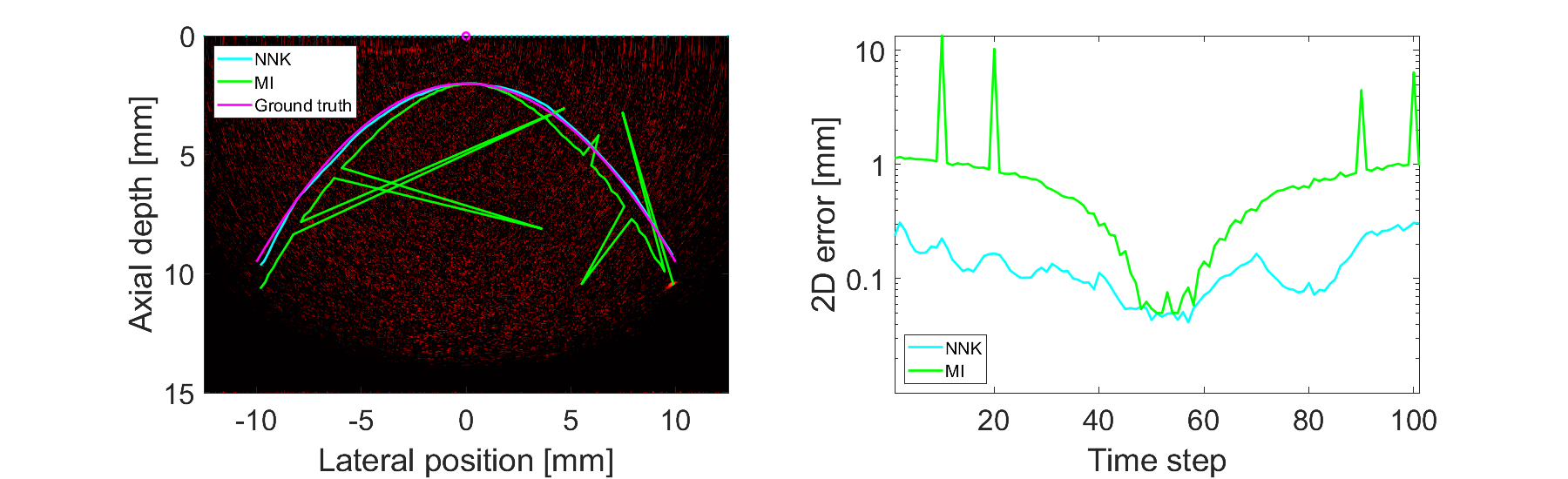}
    \caption{Tracking for synthetic experiment 2 with maximum intensity (MI) and NNK. The axial coordinate is overestimated with MI due to the absence of axial aberration correction and the location is severely misestimated in some frames because of increased noise.}
    \label{fig:MI-nnKalman}
\end{figure*}

\subsection{Simulated data} \label{sec:simulator}
A highly efficient and accurate simulator of the OpUS imaging setup, as previously described in \cite{alles2018video,alles2020source} and based on the FOCUS ultrasound simulator \cite{mcgough2004rapid,kelly2006time}, was used to evaluate the performance of our method with synthetic data examples produced with the OpUS simulator. In total, four synthetic data sets were generated to test different properties of the tracking method. Noise amplitude was computed such that SNR for in-plane locations was 6.5 dB and decreasing SNR with elevational distance, due to decreasing signal strength.
The first data set (Exp.1) comprises 101 time points and a smooth, curved object trajectory with linear motion at constant velocity in the elevational direction to test the overall performance, see Figure \ref{fig:MI-nnKalman}. We remind that our proposed method extends the MI estimation with Kalman filtering and incorporation of elevational offset estimation and axial aberration correction. Thus, Exp. 1 shows the importance of the aberration correction. We then examine other factors, such as noise in the second data set (Exp. 2), which is the same as the first one, but with tenfold noise in every tenth measured time series to investigate the robustness of the method. The third data set  (Exp. 3) follows also the same axial-lateral trajectory as Exp. 1, but the object is positioned in-plane for all frames. The fourth data set (Exp. 4) has stationary lateral and axial coordinates with a constant change in elevation, and is meant to test the accuracy of offset estimation.

\subsubsection{Reference methods for comparison}
In addition to the proposed combination of neural network tracking with Kalman filtering (NNK), we test two other reduced models: plain Gaussian random walk (NNK-RW) and independent subsequent states (NNK-I). Mathematically they differ with respect to the dynamic model: NNK-RW assumes that $\bs{x}_k = \bs{x}_{k-1} + \bs{c}_k$ and NNK-I that $\bs{x}_k = \bs{c}_k$ (compare to Eq. \eqref{eq:dynamicmodel}). 
We compared our method to MI tracking which estimates the object location as the pixel with highest intensity and thus only outputs a 2D location. To evaluate the performance of all considered methods, we computed the mean 2D Euclidean distance from the estimated axial and lateral coordinates to the ground truth using synthetic data. We additionally evaluate accuracy of the three-dimensional positional estimate with Exp. 4.  Finally, we examined the localisation accuracy of NNK as a function of depth (axial coordinate) and out-of-plane offset with an axial line trajectory simulated with different values for out-of-plane offsets. This evaluation was done using 3D Euclidean distance.
\subsection{Experimental data}
To test the out-of-plane tracking abilities of the method we \rev{performed one physical experiment closely matching simulated Exp. 4 and one to test accuracy when moving further out-of-plane. In the first experiment,} the tip of a metal pushpin (tip diameter: 50 \textmu m) was used to emulate a point object and was submerged in water as a homogeneous background medium. This pin was placed centrally within the imaging aperture at an axial distance of 7.5 mm, and was attached to a manual translation stage (PT1/M, Thorlabs, Germany) to allow for controlled motion orthogonal to the image plane (i.e.,  ``out-of-plane'') and provide ground-truth positions for quantitative evaluation. The tip of this pin was placed at out-of-plane positions ranging between -3 mm to +5 mm at a regular step size of 100 \textmu m, and at each position a 2D OpUS image was acquired. \rev{For the second physical experiment, the out-of-plane and lateral positions were varied simultaneously to mimic a non-orthogonal drift of the object. The object was initially located centrally in the image at an axial depth of 7.5 mm, and moved in increments of 100 µm (lateral) and 200 µm (out-of-plane) to a total out-of-plane position of 10 mm.} SNR for the experimental data is estimated to be around 4 dB.

\subsection{Tuning parameter selection}
The tracking algorithm requires the selection of four process noise variance parameters that can be used to fine-tune the process. Values that are too low ($< (10^{-4} \ \text{mm})^2$) may cause the estimated trajectory to be too restricted in case of rapid changes in the position or velocity of the object. With ideal data (high SNR), values that are too high have little effect, but with noisier data robustness suffers. This transition starts to take place at around the value of $(0.2 \ \text{mm})^2$. Thus, the parameters were chosen empirically as $(0.005 \ \text{mm})^2$ to allow enough flexibility to recover from sudden changes in the position and velocity but at the same time provide robustness against noise.

\begin{table*}[!th]
\centering

 \begin{tabular}{l | c | c | c | c}

Test data & NNK & NNK-RW & NNK-I  & MI \rev{(reference)}\\ 
\hline
Exp. 1: Regular & 0.12 (0.076) [0.34] & 0.12 (0.078) [0.37] & 0.13 (0.078) [0.38] & 0.62 (0.36) [1.17]\\ 
Exp. 2: Increased noise & 0.13 (0.080) [0.42] & 0.15 (0.12) [0.78] & 0.51 (1.77) [13.09] & 1.06 (2.19) [16.51] \\
Exp. 3: No offset & 0.11 (0.053) [0.26] & 0.11 (0.052) [0.26] & 0.11 (0.052) [0.26] & 0.10 (0.058) [0.28]\\
Exp. 4: Stationary axial \& lateral & 0.096 (0.040) [0.19] & 0.094 (0.035) [0.18] & 0.098 (0.039) [0.20] & 0.33 (0.34) [1.30]
\end{tabular}
\caption{\label{tab:errors}2D errors (axial/lateral) as mean distance (standard deviation) [maximum distance] in millimetres with respect to ground truth of different tracking schemes for the synthetic data experiments: proposed method (NNK), \rev{the two reduced models using Gaussian random walk (NNK-RW) and independent subsequent states (NNK-I), and the reference method} maximum intensity (MI).}
\end{table*}

\begin{figure}[!th]
    \centering
    \includegraphics[width=.5\textwidth]{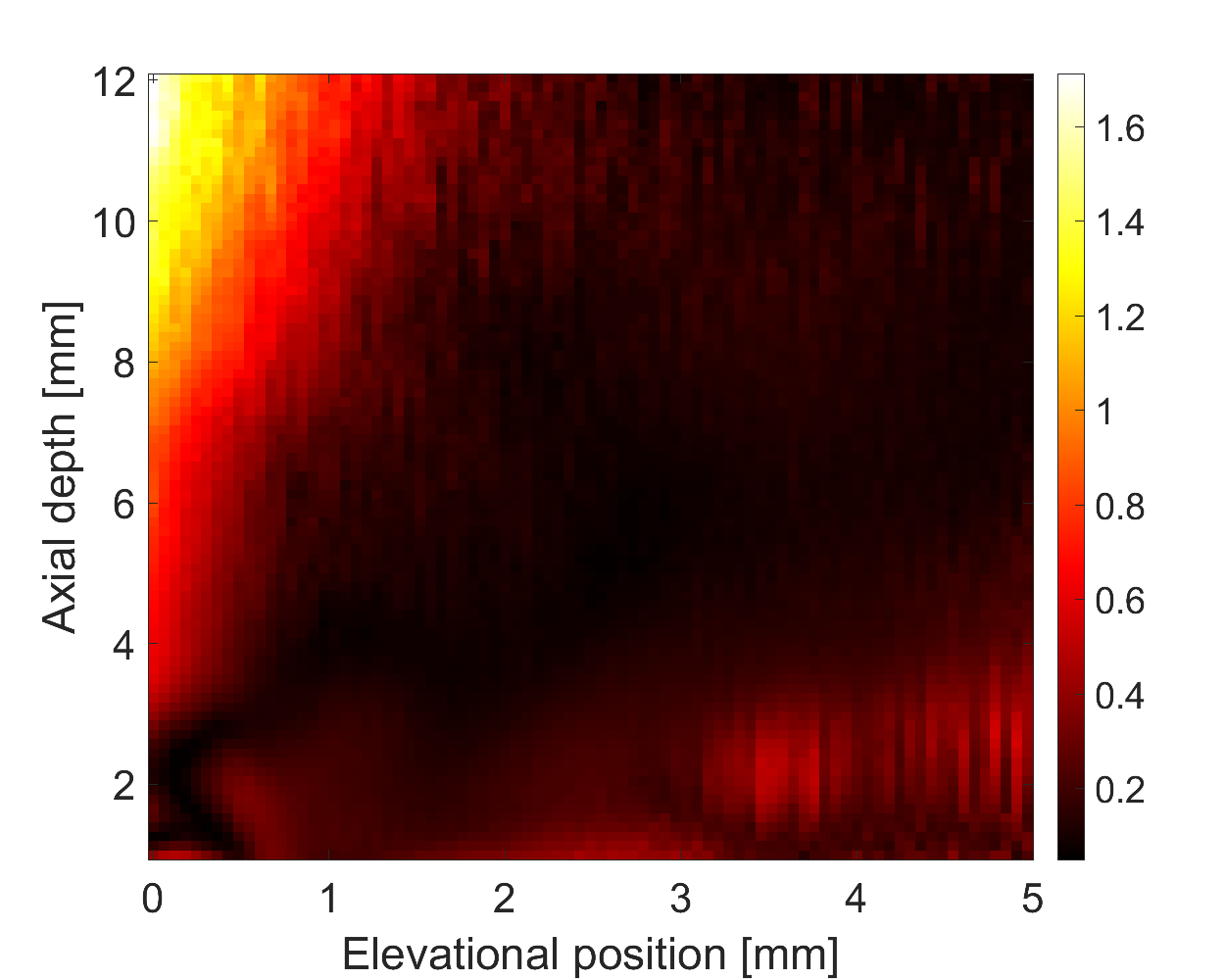}
    \caption{3D localisation error (mm) of NNK as a function of elevational position and axial depth, \rev{for lateral position at $0$~mm.}}
    \label{fig:errormatrix}
\end{figure}



\begin{figure*}[!t]
    \centering
    \includegraphics[width=\textwidth]{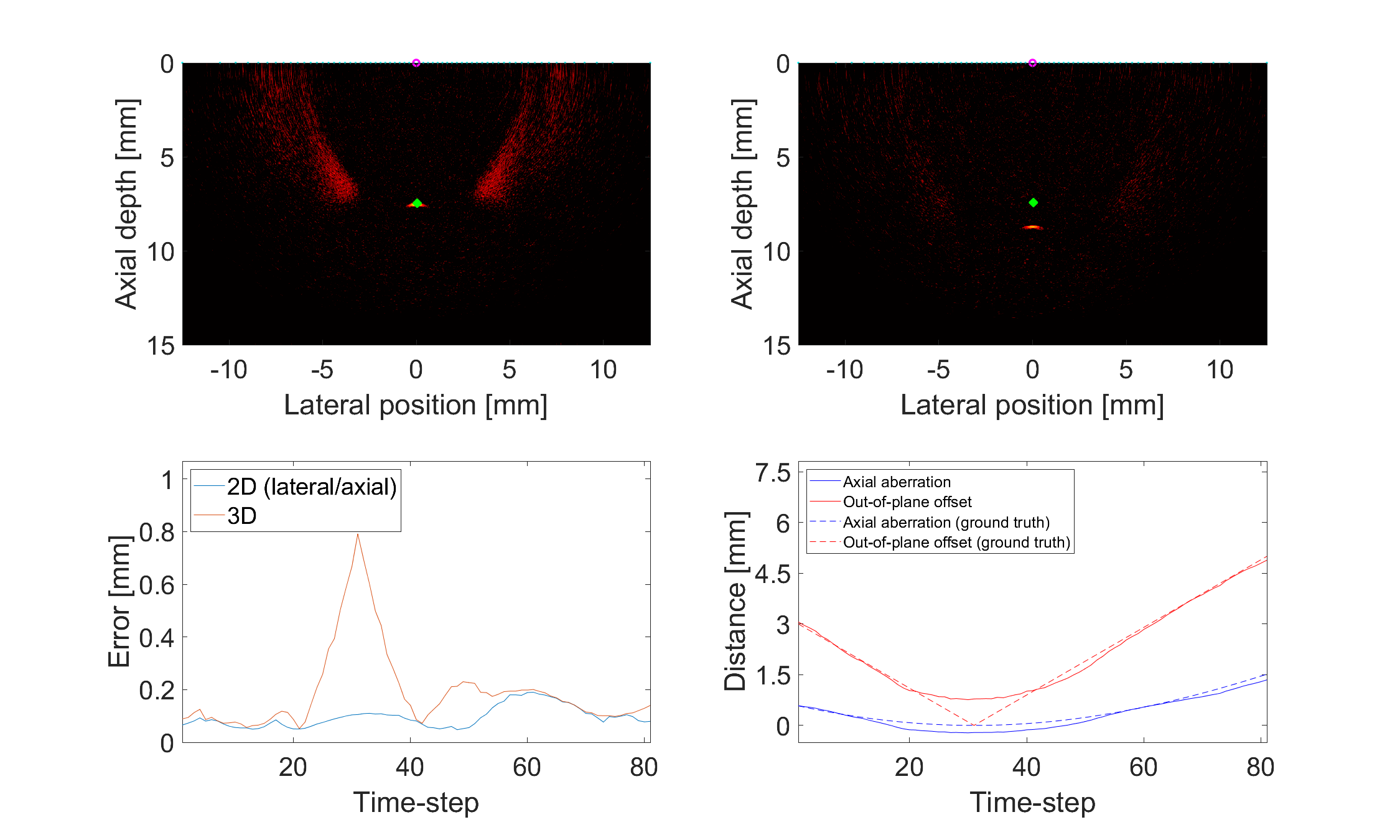} 
    \caption{Tracking for synthetic experiment 4. (Top left) tracked location (green dot) at time step 40 (out-of-plane distance $\sim$1 mm), (top right) tracked location at the last time step (out-of-plane distance 5 mm), (bottom left) 2D and 3D error (Euclidean distance) from the ground truth and (bottom right) out-of-plane offset (elevational distance) and axial aberration over time.}
    \label{fig:evosim}
\end{figure*}

\begin{figure*}[!t]
    \centering
    \includegraphics[width=\textwidth]{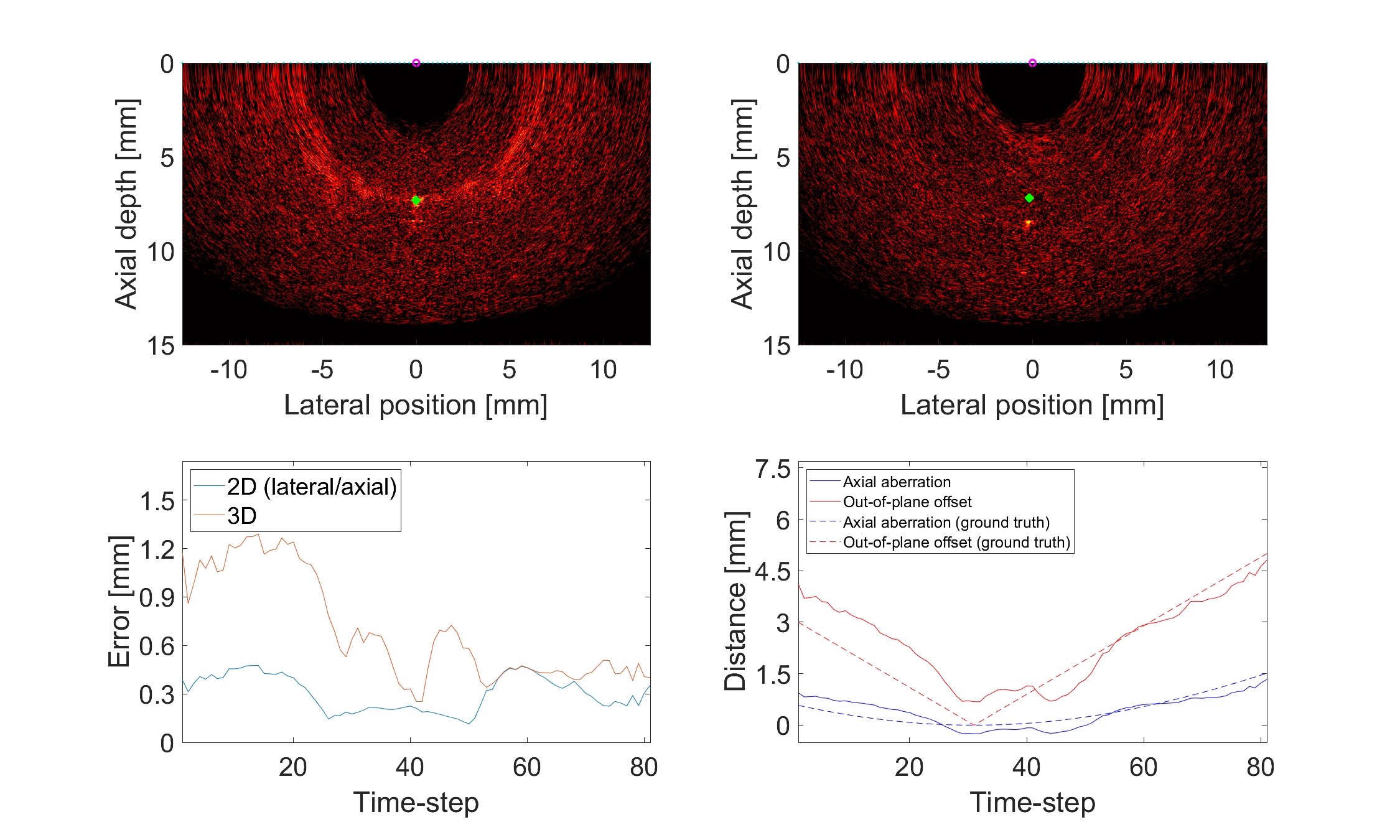} 
    \caption{Tracking for experimental dataset 1 matched to synthetic experiment 4. (Top left) tracked location (green dot) at time step 40 (out-of-plane distance $\sim$1 mm), (top right) tracked location at the last time step (out-of-plane distance 5 mm), (bottom left) 2D and 3D error (Euclidean distance) from the ground truth and (bottom right) out-of-plane offset (elevational distance) and axial image aberration over time.}
    \label{fig:evoexp}
\end{figure*}

\begin{figure*}[!t]
    \centering
    \includegraphics[width=\textwidth]{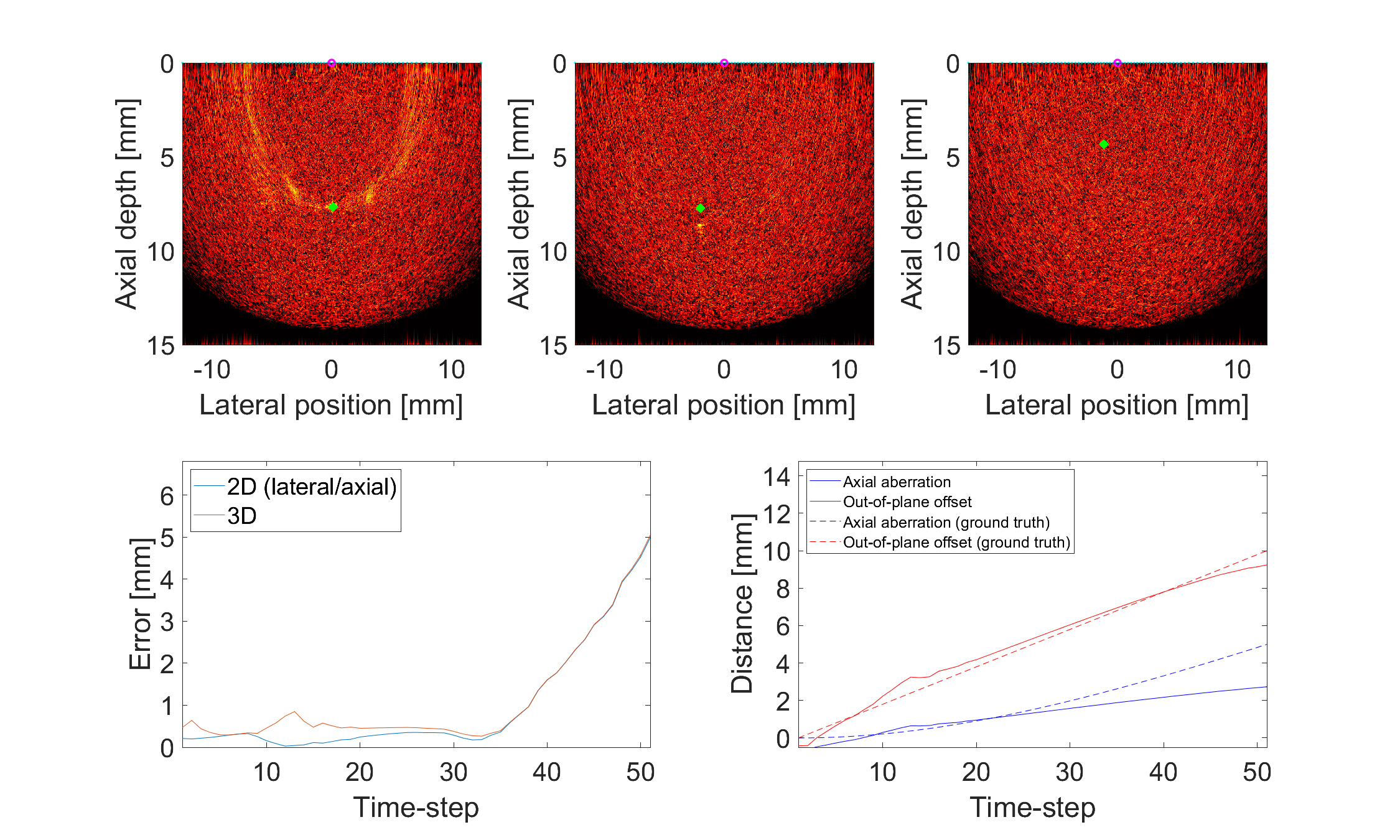} 
    \caption{\rev{Tracking for experimental dataset 2. (Top left) tracked location (green dot) at time step 1 (out-of-plane distance 0 mm), (top middle) tracked location at time step 20 (out-of-plane distance $\sim$4 mm), (top right) tracked location at the last time step (out-of-plane distance 10 mm), (bottom left) 2D and 3D error (Euclidean distance) from the ground truth and (bottom right) out-of-plane offset (elevational distance) and axial image aberration over time.}}
    \label{fig:evoexp2}
\end{figure*}

\section{Results}
\subsection{Results on simulated data}
Table \ref{tab:errors} shows the errors for the four tracking experiments and different varying methods. The proposed NNK methods perform clearly better than MI with data where the tracked object is out-of-plane, due to the correction of the axial aberration caused by out-of-plane offset: the localisation error for all NNK methods is 0.13 mm, while for MI it is fivefold. For occasionally noisier data filtering-based NNK and NNK-RW retain their performance and clearly outperform NNK-I and MI that do not assume dependence between subsequent positions, this indicates that a filtering approach is necessary to provide robustness. This benefit of filtering and axial aberration correction is clearly visible in Fig. \ref{fig:MI-nnKalman}, where MI overestimates the axial coordinate and for some noisy images the estimate jumps off the trajectory (green spikes). Nevertheless, if we consider no out-of-plane offset without additional noise, all methods perform comparably well with localisation error around 0.11 mm. In terms of worst-case performance, NNK performs the best with maximum error less than three times the mean error in every experiment. In experiment 2 with increased noise, this ratio increases to almost five with NNK-RW and to over 10 with NNK-I and MI. Videos of tracking results with NNK for experiments 1 and 4 are presented in the supplementary material (Supplementary Videos 1 and 2).

Fig. \ref{fig:errormatrix} shows how the 3D error depends on depth and out-of-plane offset \rev{for lateral position at $0$~mm}. Interestingly the error is largest ($\sim$1.6 mm) when offset is small and depth is large. For bigger offsets the error gets smaller. This indicates that there are no strong enough markers in the data to reliably estimate all three coordinates when the out-of-plane offset is small. However, the reliability increases with higher out-of-plane offsets. \rev{The relatively poor performance at shallow depths and large elevational offset (bottom right of Fig. \ref{fig:errormatrix}) is caused by the directivity of the ultrasound sources and a large propagation distance, which result in poor SNR and hence poor amplitude estimates. The same pattern was obtained for different lateral positions (data not shown), with minor fluctuations in the stable region and degradation for the extreme points close to the imaging boundaries.} 
We also note, that even though the out-of-plane offset estimation is not correct for all instances, the estimated axial aberration and filtering approach still provides accurate results, as can be seen in Fig. \ref{fig:evosim}: lateral/axial trajectory is very close to the ground truth.
\subsection{Results on experimental data}
Before we could apply NNK for tracking, the experimental data required normalisation to match the amplitude (in arbitrary units) of synthetic data. While the experimental data is clearly noisier than synthetic data, the tracking method performs reasonably well. The axial aberration correction works and out-of-plane offset largely follows the expected trajectory (see Fig. \ref{fig:evoexp} and \rev{Fig. \ref{fig:evoexp2}}). \rev{However, when going farther than 6 mm away from the imaging plane in the second experiment, the algorithm breaks down and the localisation error increases rapidly.} The axial/lateral localisation error is mostly below 0.3 mm and with mean around 0.2 mm \rev{for the first experimental dataset. For the second dataset the same holds for frames 1 to 35, after which the error starts to increase. In the first experimental dataset} most of the 3D error originates from the elevational component. \rev{This effect is not as pronounced in the second dataset.} We note that \rev{in the first dataset} the estimation accuracy is worse for the first part with negative elevational distance. This indicates that the imaging probe suffers from a source of asymmetry (e.g., acoustic shadowing by an edge, or unexpected source or receiver directivity) that has not been accurately accounted for in the numerical model. This effect can also be observed in the video of this tracking experiment in the supplementary data (Supplementary Video 3).

\subsection{Computation times}
Performing one iteration of tracking took on average 298 ms. The time was split as follows: Reconstructing the image (269 ms), finding high intensity pixels (21 ms), neural network prediction (6.6 ms), Kalman filtering (0.43 ms), and displaying the image (83 ms). Hence reconstructing the image is clearly the most time consuming task and the NNK framework only adds a small computational overhead. 

Preliminary tasks include generating training data and training the neural network. Training data with 8800 rows was generated in about an hour. Training the neural network took on average only 20 seconds with a median of 17 seconds (over 10 training attempts). Generating the training data and training the neural network has to be done only once which means that tracking is essentially performed in real-time.
Computations were performed on a workstation with AMD Ryzen Threadripper 2950X processor and 32 GB RAM. The codes for NNK are written in MATLAB while the OpUS simulator uses routines compiled from C++ for CPU.

\section{Discussion}

\subsection{Discussion of markers}
Our experiments show that magnitude of the simulated measurement data coupled with axial and lateral position is correlated with the out-of-plane offset and an axial positional aberration, as illustrated in Fig. \ref{fig:amplitude}. This correlation can be exploited with machine learning to find a nonlinear relationship between these quantities. We also found that this correlation holds with experimental data after data normalisation. Nevertheless, the tracking with experimental data is less stable and shows reduced accuracy. This can be partly attributed to reduced SNR in the measurement data as well as deviations from the ideal assumptions in the simulation, but the Kalman filtering offers a framework to partly mitigate these negative effects and is still able to provide a stable estimation of the axial/lateral coordinate.

In this work we have used an OpUS simulator and computed all markers, offset and axial aberration, from the simulated data. We note that under the assumption of a homogenous medium, we can alternatively calculate the axial aberration analytically using the point-spread function of the imaging system. Nevertheless, we have observed in conducted experiments that an analytic calculation can help for small distances in the simulated data, but will lead to decreased accuracy for the experimental datasets. Thus, computing the axial aberration from the reconstructed US images for training seems to provide more generalisable markers for the estimation process. Furthermore, this fully simulated framework can be extended to heterogeneous media.


\subsection{Offset accuracy and range}

The quantitative analysis shows that tracking accuracy is worse for small offsets and larger depths. 
Most of this error seems to be caused by the out-of-plane offset estimation, while axial and lateral components are tracked well. This indicates that even though the offset might be incorrectly estimated, the proposed axial aberration correction still works.
We attribute the difficulty of estimating small elevations to the shallow slope of the out-of-plane amplitude decay for larger axial depths, as shown in Figure \ref{fig:amplitude}. This decrease in accuracy is also seen in the error matrix in Figure \ref{fig:errormatrix}, and worsens with increasing axial distance. 
Thus, it is important to provide both offset and axial aberration, to provide accurate tracking results within the Kalman filtering. For the simulated data this effect shows symmetrically at roughly out-of-plane distance under 1 mm. For the experimental data the threshold distance is similar, but an asymmetric behaviour can be observed, where positive elevational distance is underestimated and negative overestimated, as seen in Fig. \ref{fig:evoexp}. This indicates that the purely simulated framework can in principle be transferred to the experimental case, but small asymmetries in the imaging probe would need to be investigated and accounted for to further improve the results.

The imaging system in this work uses unfocussed, weakly directional circular optical ultrasound sources that insonify a wide elevational range. Consequently, out-of-plane tracking can be performed over a large elevational range limited by the SNR of the B-scan; in this work  up to \rev{$6$~mm} for positive elevational distance. \rev{For larger out-of-plane offsets, the RF data SNR is insufficient to reliably detect the pulse-echo signal.} However, for imaging probes comprising directional sources, or in the presence of an acoustic lens, this range could be different.

\subsection{Limitations and clinical applicability}\label{sec:clinical}
In this study we show that one can successfully use the correlation between out-of-plane amplitude decay and  axial/lateral positions to estimate 3D locations from linear array data. Nevertheless, this correlation was observed in a simplified simulated and experimental setting assuming homogeneous media, i.e., a water bath in the experimental setup. In order to move towards clinically realistic scenarios we need to consider various deviations from the ideal case. \rev{In the following we discuss limitations and extensions needed for clinical applicability.}

\subsubsection{Speckle}
\rev{The tracking presented is based on maximum intensity pixels, as such} speckle of low to moderate intensity (compared to the intensity of the image of the object; for instance in the case of a highly echogenic needle tip) is not expected to interfere with the estimation procedure. However, strong speckle could result in tracking errors if only the amplitude is used as marker $\mu$. In this case, the NN would likely need to be adapted to not only extract amplitude information, but also its variation across the imaging aperture -- as this spatial variation for the actual object would differ from that of speckle signal.

\subsubsection{Inhomogeneous media}
In this work,  we have presented results for homogeneous media. \rev{For an application to inhomogeneous media with spatially varying speed of sound the NN needs to be trained differently.}
Here approximate synthetic training data could, for instance, be generated using ensemble-mean speed of sound maps observed over a group of patients\rev{, and simulated with advanced methods such as the k-Wave toolbox \cite{treeby2010k}.} In addition, acoustical attenuation would affect the extracted parameter $\mu$ and hence complicate accurate out-of-plane tracking. For applications to actual tissue, this attenuation should be included in the model used to train the NN.

\subsubsection{Object geometry}
The results presented in this work were obtained for point-like objects, such as clinically encountered in the form of microbubbles, fiducial markers, brachytherapy seeds and radio-opaque markers on surgical instruments. \rev{In order to extend the method to finite-sized objects, their ultrasound response needs to be accurately modelled. Conceptually, the method applies to} finite-sized spherically symmetric objects, such as large needle tips or spherical implants. 
However, more complicated object geometries, such as long needles or asymmetric beads, are complicated due to nonlinearities arising from high echogenicity, and ambiguities in differentiating between needle tips and shafts or different object orientations. Such objects would require further refinement in the NN markers and the underlying acoustical model to make accurate predictions.

\subsubsection{Tracking range and accuracy} 
\rev{In the experimental results presented here, an out-of-plane tracking range of up to $\pm$6 mm was demonstrated, and was limited by SNR. This range could be further extended, provided the imaging probe emits a sufficiently diverging field in the elevational direction and SNR is improved, for instance using coded excitation schemes. The optical ultrasound imaging system considered here does not apply acoustic focussing in the elevational direction, and hence is ideally suited to tracking across a wide out-of-plane range. The achieved range of $\pm$6 mm is clinically highly relevant, as correcting object placement over larger distances is typically not possible without removal and re-entry of a surgical tool. For clinical imaging systems, which typically apply elevational focussing, geometrical distortion and signal amplitude decay resulting from out-of-plane offsets will still occur, and the proposed method can still be applied, provided it is retrained. However, the out-of-plane tracking range and accuracy will depend on the tightness of the elevational focusing, and hence will vary with both the F-number of the elevational focusing lens and the axial position of the object relative to the focal distance.}

\subsubsection{Experimental setup}
\rev{Here, we used a prototype optical ultrasound imaging setup to perform experimental validation measurements. While these were reasonably successful (\textit{cf.} Figs. \ref{fig:evoexp}, \ref{fig:evoexp2}) due to the availability of a highly accurate and efficient numerical model, the developmental nature of this system limited its practicality. Slow fluctuations were observed in the efficiency of the optical ultrasound sources and the sensitivity of the detector, which resulted in unforeseen variations in the ultrasound amplitudes. As the NN estimation requires the amplitude to be accurately known, these fluctuations limited the range of object trajectories to those that could be traversed quickly. This also resulted in slight differences between simulated and experimental data. Nevertheless, the estimation network generalised well to the experimental data, and the filtering approach further stabilised the estimation process. However, in principle any ultrasound imaging system that grants access to RF data could be used, even those generating focused transmissions – although the NN would need to be re-trained for each considered setup and tracking accuracy and range will vary.}

\subsection{Extensions}
The presented framework can be extended to tracking multiple point sources. In that case, a data association task would have to be solved \cite{bayesfilter,reid1979algorithm}. This means determining which pixels belong to which target. Furthermore, this would also require a more complicated setup for training data generation and extraction of markers from the measured time series.


Another interesting avenue to pursue would be the extension to needle tracking (as opposed to point object tracking) where the shaft can be mistaken for the tip, which would require a shape detection instead of a simple tracking. Alternatively, one can overcome the shaft problem by adjusting our framework to data obtained with an active listening needle \cite{ultrasoundtracking} that relies on the reception of ultrasound pulses by a fibre-optic hydrophone (FOH) integrated into the needle.

Finally, due to the limitations mentioned in Section \ref{sec:clinical} it might be promising to consider the full RF time series as input to a convolutional neural network that is also capable of extracting geometric markers from the data. This information can be still paired with manually extracted markers, such as amplitude, to improve the tracking accuracy and robustness for future applications.

\section{Conclusion}
This work proposes a neural network and Kalman filtering approach to perform accurate and robust object tracking in 3D from linear array data. The essential step is that a neural network estimates the third dimension and its impact on the 2D US image, in form of aberration in the axial coordinate. Then Kalman filtering is performed for all coordinates to provide a robust estimate with respect to noise. We have shown that the framework can provide high accuracy in estimating axial and lateral coordinates for objects that are not in-plane as well as the corresponding elevational distance. If the point-source is too close to the imaging plane, it remains difficult to provide an accurate estimate on the elevational distance, but the proposed NNK framework is still capable to provide a robust and accurate estimate on the lateral/axial coordinate.

\section*{Acknowledgements}
We thank the three anonymous reviewers for their constructive comments which significantly helped us to improve the presentation of our work. Codes for NNK will be made available after peer review.

\bibliographystyle{IEEEtran}
\bibliography{abcbib, needleTrackingRefs}

\end{document}